\documentclass[]{aa} 
\usepackage{graphicx}
\usepackage{epsfig}
\usepackage{natbib}
\bibpunct{(}{)}{;}{a}{}{,}

\begin{document}


\titlerunning{A Hot Jupiter orbiting HD\,73256}

\title{The CORALIE survey for southern extra-solar planets.}
\subtitle{X. A Hot Jupiter orbiting HD\,73256\thanks{Based on
    observations collected with the {\footnotesize CORALIE} echelle
    spectrograph on the 1.2-m Euler Swiss telescope and the
    Str\"omgren Automatic Telescope (SAT) at La\,Silla Observatory,
    ESO Chile}}

\author{S.~Udry\inst{1} 
   \and M.~Mayor\inst{1} 
   \and J.V.~Clausen\inst{2} 
   \and L.M. Freyhammer\inst{3}
   \and B.E.~Helt\inst{2} 
   \and C.~Lovis\inst{1}
   \and D.~Naef\inst{1} 
   \and E.H.~Olsen\inst{2} 
   \and F.~Pepe\inst{1}
   \and D.~Queloz\inst{1} 
   \and N.C.~Santos\inst{1,4}
}

\offprints{stephane.udry@obs.unige.ch}

\institute{Observatoire de Gen\`eve, 51 ch. des Maillettes, CH-1290
  Sauverny, Switzerland 
\and 
Niels Bohr Institute for Astronomy, Physics and Geophysics;
Astronomical Observatory, Juliane Maries Vej~30, DK-2100 Copenhagen
{\O}, Denmark 
\and 
Royal Observatory of Belgium, Ringlaan 3, 1180 Brussel and University
of Brussels (VUB), Pleinlaan 2, 1050 Brussels, Belgium
\and 
Centro de Astronomia e Astrof{\'\i}sica da Universidade de Lisboa,
Observat\'orio Astron\'omico de Lisboa, Tapada da Ajuda, 1349-018
Lisboa, Portugal }

\date{Received / Accepted } 

\abstract{ Recent radial-velocity measurements obtained with the
  {\footnotesize CORALIE} spectrograph on the 1.2-m Euler Swiss
  telescope at La~Silla unveil the presence of a new Jovian-mass Hot
  Jupiter around {\footnotesize HD}\,73256. The 1.85-M$_{\rm Jup}$
  planet moves on an extremely short-period ($P$\,=\,2.5486\,d),
  quasi-circular orbit. The best Keplerian orbital solution is
  presented together with an unsuccessful photometric
  planetary-transit search performed with the SAT Danish telescope at
  La~Silla. Over the time span of the observations, the photometric
  follow-up of the candidate has nevertheless revealed a
  $P$\,$\simeq$\,14-d photometric periodicity corresponding to the
  rotational period of the star. This variation as well as the
  radial-velocity jitter around the Keplerian solution are shown to be
  related to the fair activity level known for {\footnotesize
    HD}\,73256.
  
   }

\maketitle

\section{Introduction}

The increasing timebase of the radial-velocity surveys searching for
extra-solar planets allows the different planet-hunter teams to
announce new planetary candidates on longer and longer periods.  This
growing period-interval coverage is very important. With the
enlargement of the available statistics, new properties of the
planetary period distribution are emerging.  They provide interesting
new constraints for the migration scenarios \citep[for a recent review
see][]{Udry-2003:d}.

On the other hand short-period planets, easier to detect because of
the larger reflex motions induced on the primaries and the better
phase coverage for orbital-element determinations, were rapidly
detected. An almost complete census of these systems is available in
well-covered surveys. In the case of our {\footnotesize CORALIE}
planet-search programme \citep{Udry-2000:a}, the number of targets is
very large ($\sim$\,1650) and there are still candidates with only few
measurements. From this survey, we present here a new short-period
planet orbiting the star {\footnotesize HD}\,73256, detected thanks to
very recent radial-velocity observations.

The parent star description, radial-velocity measurements, orbital
solutions and inferred planetary characteristics for this new
candidate are presented in Sects.~2 and 3.
Hot Jupiters are promising candidates for photometric transit searches
so, in Sect.~4, we describe our photometric observations aiming first
to detect a potential planetary transit and then to check the star
photometric stability.  In Sect.~5, we discuss photometric and line
bisector measurements in relation to the activity-induced
radial-velocity jitter observed on top of the Keplerian orbital
variation.  Finally, Sect.~6 gives a summary of the results and some
concluding remarks.

\section{Stellar characteristics of HD\,73256}

{\footnotesize HD}\,73256 ({\small HIP}\,42214) is a G8/K0 dwarf in
the southern {\sl Pyxis} constellation.  The {\footnotesize HIPPARCOS}
catalogue \citep{ESA-97} lists a visual magnitude $V$\,=\,8.08, a
colour index $B$--$V$\,=\,0.782, and a precise astrometric parallax
$\pi$\,=\,27.38\,$\pm$\,0.77\,mas that sets the star at a distance of
36.5\,pc from the Sun. Its absolute magnitude is then estimated to be
$M_V$\,=\,5.27, slightly overluminous for a typical G8 dwarf. This is
probably due to the enhanced metallicity content of the star (see
below).

From a high-resolution spectroscopic abundance study of {\footnotesize
  HD}\,73256, we have determined precise values of its effective
temperature ($T_{\rm eff}$\,=\,5570\,$\pm$\,40\,K), metallicity
([Fe/H]\,=\,0.29\,$\pm$\,0.05) and gravity
($\log{g}$\,=\,4.66\,$\pm$\,0.10), using a standard local
thermodynamical equilibrium (LTE) analysis \citep[see][ for error
estimates]{Santos-2000:c}. From calibrations of the width and surface
of the {\footnotesize CORALIE} cross-correlation functions \citep[CCF;
described in][]{Santos-2002} we can also derive estimates of
$v\sin{i}$\,=\,3.22\,km\,s$^{-1}$ and [Fe/H]\,=\,0.27\footnote{These
  calibrated estimates, although {\sl statistical}, are reliable. A
  comparison between spectroscopic and CCF-calibrated metallicities
  yields a small uncertainty of only 0.05\,dex due to the calibration
  \citep{Santos-2002}}. The high metal content is a recurrent property
of stars hosting planets \citep[for a review see e.g.][ and references
therein]{Santos-2003:a}.
 
From the colour index and $T_{\rm eff}$ we derive a bolometric
correction $BC$\,=\,$-0.122$ \citep{Flower-96}.
The star luminosity is then estimated to be $L$\,=\,0.69\,L$_\odot$.
Note however that Flower's calibrations do not take metallicity into
account.  According to the tracks of the Geneva evolution models with
appropriate metal abundance \citep{Schaerer-93}, the position of the
star in the HR diagram indicates a mass
$M_\star$\,$\simeq$\,1.05\,M$_{\odot}$.  This mass is higher than
typical values for G8/K0 dwarfs because of the high metallicity of the
star.

\begin{table}[t!]
\caption{
\label{tab1}
Observed and inferred stellar parameters for 
{\footnotesize HD}\,73256. Definitions and sources 
of the quoted values are given in the text.}
\begin{tabular}{lcl}
\hline
\multicolumn{1}{l}{Spectral Type} & G8/K0 & \\
V     & 8.08   & \\
$B-V$ & 0.782  & \\
$\pi$ & $27.38 \pm 0.77$ &[mas] \\
$M_V$ & 5.27   & \\
$L$   &0.69    &$[L_\odot]$ \\
$T_{\rm eff}$ &$5570 \pm 50$   & [K] \\
$\log{g}$     &$4.66 \pm 0.10$ & [cgs] \\
$[Fe/H]$ &$0.29 \pm 0.05$ & \\
$M_\star$ &1.05  &$[M_\odot]$ \\ 
$R_\star$ &0.89  &$[R_\odot]$ \\
$v\sin i$ &$3.22 \pm 0.32$ &[km\,s$^{-1}$] \\
$\log(R^{\prime}_{HK})$ & $-4.49$ & \\
age($R^{\prime}_{HK}$) &0.830 &[Gyr] \\
$P_{\rm rot}(R^{\prime}_{HK})$\hspace{0.25cm} 
        &13.90 & [days]  \\
$P_{\rm rot}$(phot)\hspace{0.25cm} 
        &13.97 & [days]  \\
$v_{\rm eq}$ &3.26 &[km\,s$^{-1}$] \\
\hline
\end{tabular}
\end{table}

\begin{figure}[t!]
\psfig{width=\hsize,file=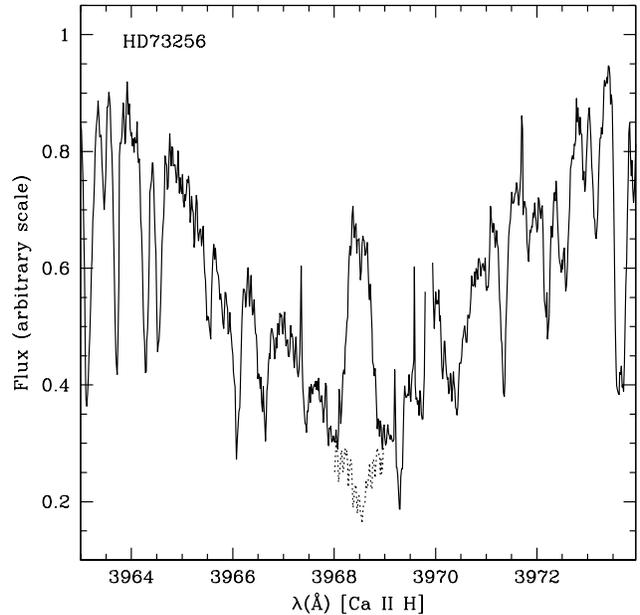}
\caption{
\label{fig1}
\mbox{$\lambda$ 3968.5 \AA\ \ion{Ca}{ii}\,H} absorption line region of
the summed {\small CORALIE} spectra for {\footnotesize HD}\,73256.  A
clear emission feature is observed. The dotted line in the center of
the line represents the spectrum of a non-active star.}
\end{figure}

The models also suggest a completely unconstrained age close to 1\,Gyr
for the star, in agreement with its measured high activity level.
{\footnotesize HD}\,73256 belongs to the sample surveyed by
\citet{Henry-96} for \ion{Ca}{ii}\,H and K chromospheric emission. It
is found to be fairly active with an index
$\log{R^{\prime}_{HK}}$\,=\,$-4.49$. Chromospheric emission is also
directly visible on the coaddition of our {\small CORALIE} spectra
(Fig\,\ref{fig1}).  Following the calibration by \citet{Donahue-93},
this activity index value points towards a young stellar age around
830\,Myr.

From the relation between the activity index and stellar rotation
period \citep{Noyes-84}, we derive a period of rotation $P_{\rm
  rot}$\,$\simeq$\,13.9\,days for {\footnotesize HD}\,73256.  Assuming
that the orbital and rotation axes coincide\footnote{This is verified
  for HD\,209458 \citep{Queloz-2000:a}}, a ``statistical'' equatorial
velocity $v_{\rm eq}$ can be derived from the radius of the star.  The
orbital plane inclination is then obtained from the measured projected
rotational velocity $v\sin i$\,=\,3.22\,km\,s$^{-1}$. Using the simple
relation between stellar luminosity, radius and effective temperature
$L$\,=\,$4\pi R^2\sigma T_{\rm eff}^4$ and with the stellar parameter
values given above, the radius is estimated to be
$\simeq$\,0.89\,R$_\odot$. This leads to a value $v_{\rm
  eq}$\,$\simeq$\,3.26\,km\,s$^{-1}$, very close to the quoted
$v\sin{i}$ ($\sin{i}$\,$\simeq$\,0.98).  The true mass of the planet
is thus not expected to be very different from the derived minimum
mass. The short period and favourable inclination make {\footnotesize
  HD}\,73256 a good candidate for a photometric transit search (see
Sect.\,\ref{sect4}).

The observed and inferred stellar parameters are summarized in
Table~\ref{tab1}.  Due to the fair activity level of the star some
radial-velocity jitter is expected on a typical timescale of the order
of the rotational period. This is discussed in Sect.~\ref{sect5}.

\section{{\footnotesize CORALIE} orbital solution for HD\,73256}

\begin{figure}[t!]
\psfig{width=\hsize,file=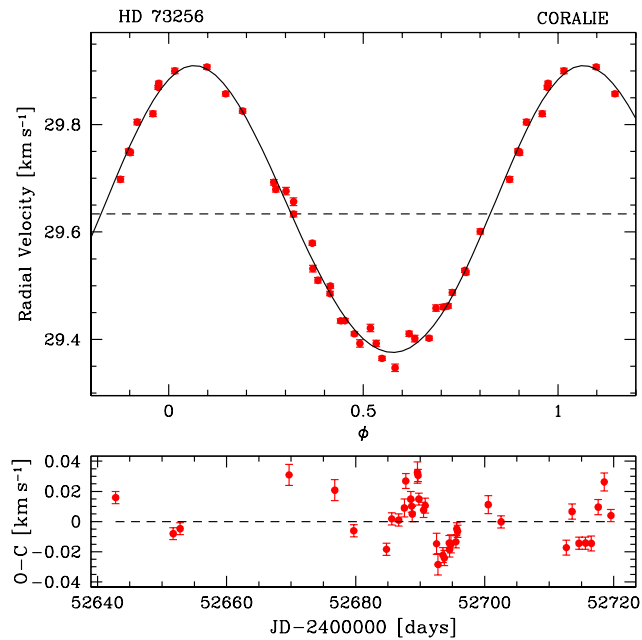}
\caption{
\label{fig2}
Phase-folded radial-velocity measurements obtained with {\footnotesize
  CORALIE} for {\footnotesize HD}\,73256 (top). Error bars represent
photon-noise errors.  The still large residuals around the solution
(bottom) show some structure and may be explained by activity-induced
jitter (see Sect.\,\ref{sect5}). For clarity of the diagram, the
measurement on JD~2\,451\,964 is not displayed in the figure.}
\end{figure}

The first observation with {\footnotesize CORALIE} of {\footnotesize
  HD}\,73256 started in February 2001 (JD\,=\,2\,451\,964.67).
However, almost 2 years elapsed before we observed the star again.
Because of the clear radial-velocity difference between the 2
measurements, the star was then intensively followed during more than
2 months. Like this, we have gathered 40 precise radial velocities.
The photon-noise errors of individual measurements are typically of
4--5\,m\,s$^{-1}$ despite the relative faintness of the star for our
1.2-m telescope.

The best Keplerian model reproducing the observations yields an
accurately constrained orbital period of 2.54858\,$\pm$\,0.00016\,days,
a non-significant eccentricity $e$\,=\,0.029\,$\pm$\,0.02, and a
semi-amplitude $K$\,=\,269\,$\pm$\,8\,m\,s$^{-1}$ of radial-velocity
variation.  Uncertainties are estimated through Monte-Carlo
simulations. The phase-folded radial-velocity curve is displayed in
Fig.~\ref{fig2} (top).

Using the derived 1.05\,M$_\odot$ mass for {\footnotesize HD}\,73256,
the best-fit parameters lead to a companion {\sl minimum} mass
$m_2\,\sin{i}$\,=\,1.87\,M$_{\rm Jup}$ and a separation
$a$\,=\,0.037\,AU between the star and the planet.  This inferred
separation is the 3$^{\rm rd}$ smallest known to date amongst hot
Jupiters, after the 2 {\footnotesize OGLE}
\citep{Udalski-2002:a,Udalski-2002:b} candidates recently proposed
\citep{Konacki-2003,Dreizler-2003} but for which stronger
radial-velocity confirmations are, however, required.  At such a small
distance the planet is strongly heated by its parent star.  From
recent models of irradiated planets with condensed dust atmospheres,
Barman et al. (in prep) estimate the planet {\sl day-side} temperature
to be around 1500\,K \citep[see also][ for intrinsic temperature
estimates]{Baraffe-2003}. The complete set of orbital elements with
their uncertainties and the inferred planetary parameters are given in
Table~\ref{tab2}.

\begin{table}[t!]
\caption{
\label{tab2}
{\footnotesize CORALIE} best Keplerian orbital solution derived for 
{\footnotesize HD}\,73256 as well as inferred planetary parameters. 
Uncertainties are estimated through Monte-Carlo simulations.}
\begin{tabular}{l@{}lr@{\hspace{0.25cm}$\pm$\hspace{0.25cm}}l}
\hline
$P$ &$[$days$]$     &2.54858   &0.00016 \\
$T$ &[JD-2400000]   &52500.18  &0.28   \\
$e$ &               &0.029     &0.02  \\
$V$ &[km\,s$^{-1}$] &29.729    &0.005  \\
$\omega$ &[deg]     &337.3     &45.8   \\
$K$ &[m\,s$^{-1}$]  &269       &8      \\
$N_{\rm meas}$ &    &\multicolumn{2}{c}{40} \\
$\sigma (O-C)$\hspace{.25cm}  & [m\,s$^{-1}$] 
                    &\multicolumn{2}{c}{14.8} \\
\hline
$a_1\sin i$ &[$\mathrm{10^{-5}}$\,AU]   &6.31 &0.16  \\  
$f(m)$ &$\mathrm{[10^{-9}\,M_{\odot}]}$ &5.15 &0.40 \\
$m_{2}\,\sin i$ &$\mathrm{[M_{\rm Jup}]}$ &1.87 &0.49 \\
$a$ &[AU] & \multicolumn{2}{c}{0.037} \\
$T_{\rm day\, side}$ &[$^\circ$K] & \multicolumn{2}{c}{1500} \\
\hline
\end{tabular}
\end{table}

The measured weighted r.m.s. around the solution is
$\sigma(O-C)$\,=\,14.8\,m\,s$^{-1}$, a high value compared to the
individual photon-noise error. Moreover, some structure is clearly
apparent in the residuals drawn as a function of the Julian date
(Fig.\,\ref{fig2}, bottom).
The possible periodicity of the residuals is discussed in
Sect.\,\ref{sect5} in relation with the activity and rotational period
of the star.

\section{Photometric observations}
\label{sect4}

With its 2.55-day period, the {\footnotesize HD}\,73256 system is a
good candidate for photometric transit search. Furthermore, as
discussed above, a favourable geometry could be expected from activity
indicator and rotational velocity considerations. We have thus rapidly
launched an intensive campaign of high-precision differential
photometry in order to detect a possible planetary transit.

\begin{figure}[t!]
\psfig{width=\hsize,file=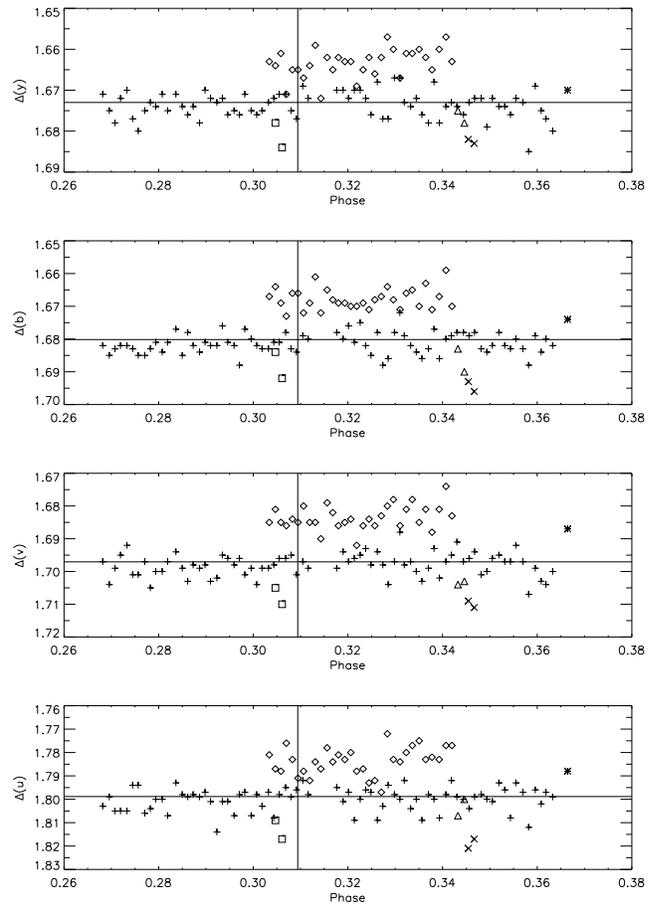}
\caption{
\label{fig3}
SAT differential photometric data ({\footnotesize HD}\,73256 --
{\footnotesize HD}\,72673, instrumental system) around the orbital
phase of the potential transit (vertical line), in four different
colours of the {\sl Str\"omgren} photometry, for different nights of
observation (different symbols). Typical r.m.s. errors of one
magnitude difference are 0.003-0.004 ($ybv$) and 0.005-0.006 ($u$).
The horizontal lines represent the mean magnitude differences (all
observations).  No indication of a transit is found.}
\end{figure}

The photometric $uvby$ observations were obtained with the {\sl
  Str\"omgren Automatic Telescope} (SAT) at ESO La\,Silla, Chile.
Details on the standard observational and reduction strategy can be
found in \citet{Clausen-2001}. {\footnotesize HD}\,72673,
{\footnotesize HD}\,72954AB, and {\footnotesize HD}\,71583 were used
as comparison stars.
Continuous differential observations were carried out, on two nights,
for several hours around the predicted transit times
(Fig.\,\ref{fig3}). Typical r.m.s. errors of one magnitude difference
are 0.003-0.004 ($ybv$) and 0.005-0.006 ($u$).  Unfortunately, no
transit indication is found. This implies an orbital inclination
smaller than $i$\,$\simeq$\,82.5\,$^{\circ}$
(i.e. $\sin{i}$\,$\leq$\,0.99).

\begin{figure}[t!]
\psfig{width=\hsize,file=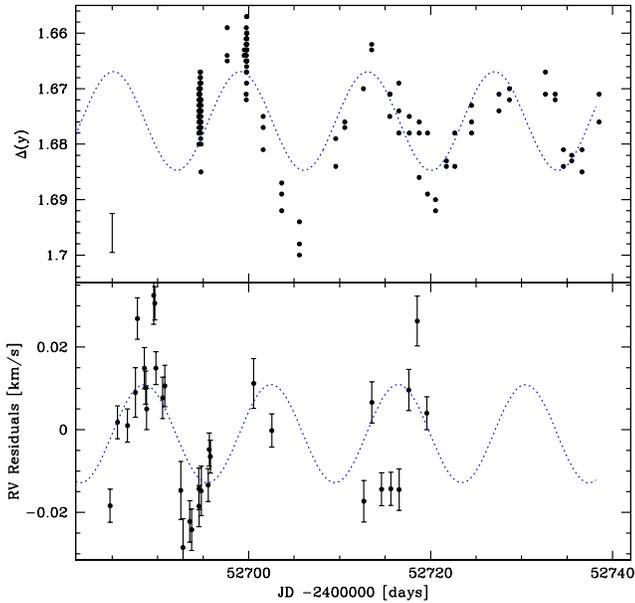}
\caption{
\label{fig4}
{\bf Top.} SAT photometric observations of {\footnotesize
  HD}\,73256\,--\,{\footnotesize HD}\,72673 in the $y$ band
(instrumental system).  The typical uncertainty on $\Delta(y)$ is
shown in the lower left corner.  A periodic structure is clearly
appearing in the data. A sine curve with a 13.97-d period (from a
Fourier transform) is adjusted to the data and displayed for
illustration.  {\bf Bottom.}  Simultaneous radial-velocity residuals
around the best Keplerian solution. A sine curve with the {\sl
  photometric} period is also displayed. The phase shift between the
two sine curves is 1.53\,rad, close to $\pi$/2, the expected value for
spot-induced photometric and radial-velocity variations
\citep{Queloz-2001:a}.}
\end{figure}

However, the data show different magnitude levels from one night to
the other (different symbols in the figure). The star is also known to
be active. Thus, in order to check for photometric variability,
{\footnotesize HD}\,73256 was also regularly monitored, over several
additional nights, to cover the complete interval of orbital phases.
The result is shown in Fig.\,\ref{fig4} (top) displaying the observed
differential magnitudes ($\Delta(y)$) as a function of the Julian
date. A periodic variation is clearly visible over the time span of
the observations. A Fourier transform of the data\footnote{Without the
  last few days for which the photometric phase seems to have changed}
yields a period of 13.97~days (Fig.\,\ref{fig5}), in very good
agreement with the rotational period derived from the activity index
(13.9~days). The photometric variation can thus be interpreted as due
to spots on the surface of the star.  Such spots not only perturb the
observed stellar luminosity over a few rotational periods but also
induce a spurious radial-velocity jitter, observable as a
radial-velocity variation of similar periodicity (see
Sect.\,\ref{sect5.2} below).

\section{Effect of stellar activity}
\label{sect5}

\subsection{Ruling out activity as the origin of the main
  radial-velocity signal}

\begin{figure}
\psfig{width=\hsize,file=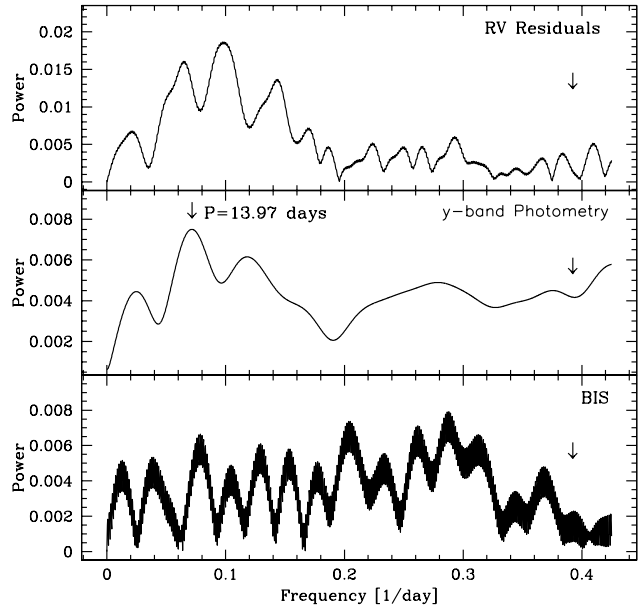}
\caption{
\label{fig5}
Fourier transforms of radial-velocity residuals around the Keplerian
solution (top), SAT photometric $\Delta(y)$ data (middle), and CCF-BIS
measurements (bottom). Photometric and radial-velocity observations
share similar behaviours in the frequency space, whereas no coherence
is found in the BIS measurements. The rotation period is found at
$P_{\rm rot}$\,=\,13.97~days (middle). The vertical arrow on the right
hand side in each panel indicates the location of the orbital period.}
\end{figure}

\begin{figure}[t!]
\psfig{width=0.85\hsize,file=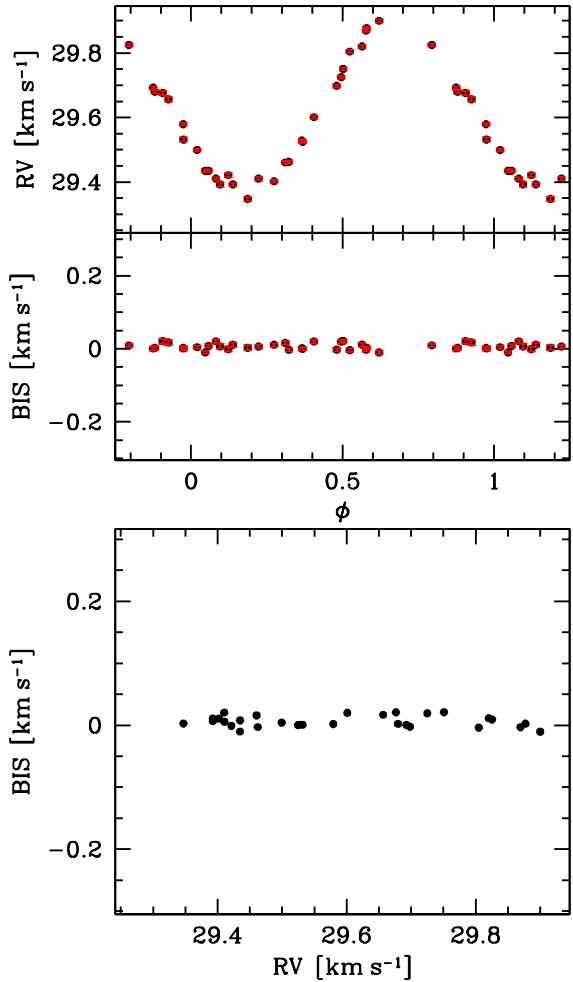}
\caption{
\label{fig6}
{\bf Top.} Radial velocities (RV, upper panel) and inverse bisector
slope (BIS, middle panel) phased with the orbital period
($P$\,=\,2.5486\,d) for {\footnotesize HD}\,73256. {\bf Bottom.}  RV
vs BIS plot (same scale) showing the independence of the two
quantities. }
\end{figure}

Most extra-solar giant planets have been detected through variations
of the observed radial-velocity of solar-type stars. Such variations
may also be induced by stellar activity related phenomena (like spots)
over a few rotational periods. The corresponding radial-velocity
amplitude can reach a few tens of m\,s$^{-1}$
\citep{Saar-97,Saar-98,Santos-2000:b}. Moreover, this amplitude
increases with the rotational velocity of the star.  For
{\footnotesize HD}\,73256, $v\sin{i}$ is not very high, so we do not
expect much effect from activity.  Nevertheless, the measured fair
activity level of the star calls for an {\sl a posteriori} check to
assert the orbital solution as the best explanation for the
observations.  First, the photometric measurements vary with a
timescale much longer than the orbital period suggesting different
origins for the radial-velocity and photometric variations. This can
be further checked by controlling the shape of the spectral lines,
expected to vary in the case of spot-induced radial-velocity
variations. A powerful diagnostic is directly available from our
spectra by computing the bisector inverse slope (BIS) of the
cross-correlation functions used for the radial-velocity determination
\citep{Queloz-2001:a}.  In case of phased variations of radial
velocities and bisector slopes, we expect a tight correlation between
these two quantities. This is not observed for the 2.55-day period
phasing of the data (Fig\,\ref{fig6}). Also, no indication of
variation of the photometric and BIS data is seen in the Fourier space
at the position of the orbital period (Fig.\,\ref{fig5}).  We thus can
exclude activity as the source of the observed main radial-velocity
variation. A planetary companion is the best explanation.

\subsection{Activity as the source of radial-velocity noise}
\label{sect5.2}
  
We have noticed in previous sections that the star presents periodic
patterns in the photometric data and in the residuals around the
Keplerian orbital solution as well.  We shall now examine the
possibility for these variations to be explained by stellar
activity-induced spots.

First, we show that the two quantities present variations of similar
timescales (13.97\,days), furthermore compatible with the rotation
period deduced from the activity index (13.9~days). This is readily
visible in Fig.\,\ref{fig4} where available simultaneous photometric
points and radial velocities (here the residuals around the Keplerian
model) are plotted as a function of the Julian date.  Fourier
transforms of the data also yield similar behaviours in the frequency
space (Fig.\,~\ref{fig5}).  In Fig.\,\ref{fig4}, illustrative sine
curves with fixed {\sl photometric} period are plotted on top of the
points. The phase shift between the two curves (1.53~rad) is close to
$\pi$/2, exactly what is predicted for photometric and radial-velocity
variations induced by activity-related spots, as shown on
{\footnotesize HD}\,166435 by \citet{Queloz-2001:a}. Very similar
results are obtained in the 4 different bands of the Str\"omgren
photometry.

\begin{figure}[t!]
\psfig{width=0.875\hsize,file=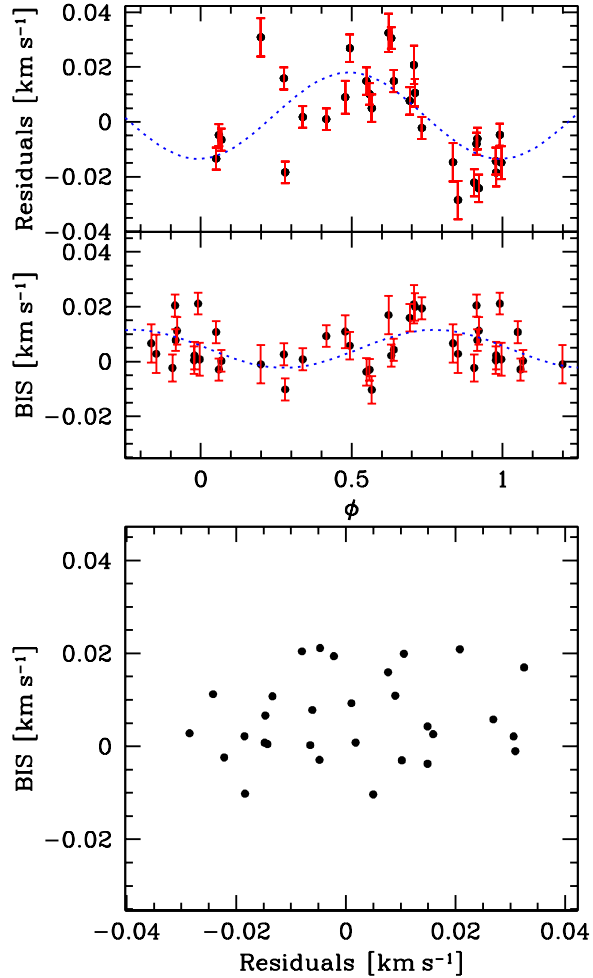}
\caption{
\label{fig7}
Same as Fig.\,\ref{fig6} but for the residuals around the Keplerian
solution, phased with the photometric period ($P$\,=\,13.97\,d).}
\end{figure}

Another approach consists in comparing the residuals around the
orbital solution to the BIS measurements, in the same way as for the
actual radial velocities.  Figure\,~\ref{fig7} (top) presents the CCF
bisector inverse slope (BIS) measurements and radial-velocity
residuals, phased with the rotational period derived from the
photometric data. Although the two quantities do not correlate clearly
(Fig.\,\ref{fig7}, bottom), some hints of coupled variations are
emphasized in the top panels by the sine curves with fixed 13.97-d
period fitted to the data. We are probably reaching here the limit of
the method for the available spectral resolution and radial-velocity
precision, especially if the BIS signature is much smaller than the
radial-velocity jitter in the case of weak rotators, as suggested by
\citet{Santos-2003:b}.

The above considerations unambiguously show the link between stellar
activity and the radial-velocity spurious noise observed around the
orbital solution.  Important progresses are expected in the domain
with the higher quality spectra and radial velocities to be provided
by {\footnotesize HARPS} \citep{Pepe-2002:b}.

\section{Summary and concluding remarks}

We have described here a new Hot Jupiter candidate orbiting the star
{\footnotesize HD}\,73256, detected with {\footnotesize CORALIE}
as part of our large planet-search survey in the southern hemisphere
\citep{Udry-2000:a}. The planet is on a quasi-circular orbit, very
close to its parent star. The period is $P$\,=\,2.5486\,days. The
inferred minimum planetary mass is 1.85\,M$_{\rm Jup}$ and the
star-planet separation is only 0.037\,AU.

A dedicated photometric search for a potential planetary transit has
been undertaken as soon as the orbital parameters were known with
suitable precision. Although unsuccessful, the photometric
measurements allowed us to estimate a stellar rotation period of
13.97~days, much longer than the orbital period. This rules out
activity as the source of the observed radial-velocity variation. A
further confirmation is brought by CCF bisector measurements showing
no correlation with the radial velocities.  The high activity level of
the star is however thought to be responsible for the somewhat large
radial-velocity residuals observed around the best Keplerian solution.
Comparable variation timescales for the photometric data and residuals
support this idea.

Concerning activity-related noise, it is worth noticing that the
planet orbits a fairly active late-type dwarf. This is also the case
for {\footnotesize HD}\,130322 \citep{Udry-2000:a} and {\footnotesize
  HD}\,192263\footnote{The existence of this planet was recently
  questioned by photometric measurements showing a periodicity
  compatible with the orbital period \citep{Henry-2002}.  New
  simultaneous photometry and radial velocities show, however, that
  the two quantities are not correlated \citep{Santos-2003:b},
  contrarily to what is expected in the case of activity (spots)
  origin for the radial-velocity variations. The planet orbiting
  HD\,192263 is thus back in the list}
\citep{Santos-2000:a,Santos-2003:b}. In these examples the
radial-velocity jitter induced by activity is small compared to the
orbital radial-velocity semi-amplitude and thus does not prevent us
from detecting the planet. This illustrates the result pointed out by
\citet{Saar-98} and \citet{Santos-2000:b} that activity-induced
radial-velocity noise is becoming smaller when going from late F to K
dwarfs, a trend mostly due to the typical lower rotation rate of the
latters. K dwarfs thus remain suitable targets for planet-search
programmes even if they show a significant activity level.

Finally, once again the combination of photometry, CCF bisector
analysis and cross-correlation technique \citep{Queloz-2001:a} has
provided the mandatory and robust diagnostics for confidently ruling
out activity as the source of the observed radial-velocity variation.

\begin{acknowledgements}
  We are grateful to the staff from the Geneva Observatory which is
  maintaining the 1.2-m Euler Swiss telescope and the CORALIE echelle
  spectrograph at La\,Silla. We thank the Geneva University and the
  Swiss NSF (FNRS) for their continuous support for this project.
  Mohamed Yacine Bouzid kindly observed HD\,73256 on a few nights at
  SAT. The SAT group gratefully acknowledges the technical staff at
  Copenhagen University for its excellent support.  The photometry was
  obtained as part of an extensive study of GK-type eclipsing binaries
  supported by the Danish Natural Science Research Council.  L.M.F.
  acknowledges support from the project IUAP P5/36 financed by the
  Belgian State. Support from Funda\c{c}\~ao para a Ci\^encia e
  Tecnologia, Portugal, to N.C.S., in the form of a scholarship is
  gratefully acknowledged.  This research has made use of the Simbad
  database, operated at CDS, Strasbourg, France
\end{acknowledgements}


\bibliographystyle{aa} 
\bibliography{udry_articles}

\end{document}